\documentclass{article} 
\usepackage[preprint]{colm2026_conference}

\usepackage{microtype}
\usepackage{graphicx}
\usepackage{booktabs}
\usepackage{amsmath}
\usepackage{amssymb}
\usepackage{xcolor}
\usepackage{multirow}
\usepackage{float}
\usepackage{placeins}
\usepackage{lineno}

\usepackage{hyperref}
\usepackage{url}
\definecolor{darkblue}{rgb}{0, 0, 0.5}
\hypersetup{colorlinks=true, citecolor=darkblue, linkcolor=darkblue, urlcolor=darkblue}

\usepackage{mathtools}
\usepackage{amsthm}

\title{Confident and Wrong: Silent Semantic Failures \\ in Coding Agents}

\author{Aman Mehta \\ Snowflake AI Research \\ \texttt{aman.mehta@snowflake.com}}

\begin{document}

\ifcolmsubmission
\linenumbers
\fi

\maketitle

\begin{abstract}
As coding agents move into production workflows, teams need to know not only whether an agent completes a task, but whether its action can be trusted. We show that completion and trustworthiness diverge sharply and systematically. Across 1{,}750 trajectories on 50 SWE-bench Verified tasks, we compare four frontier models over repeated runs and separate \emph{submit rate} from test-verified \emph{resolve rate}. GPT-5 submits a patch on 100\% of runs but resolves only 44\%; Llama~4 submits on 99\% but resolves 18\%; and Gemini, despite submitting least often at 70\%, resolves more tasks than GPT-5 (50\% versus 44\%). These gaps are not random: they concentrate in one dangerous failure mode we call \emph{silent semantic failure}. Qualitatively, on a buggy task the agent submits a plausible-looking patch on all five runs, yet none pass, the same misinterpretation repeated rather than random error. Quantitatively, it dominates failure, covering 80\% of Llama~4's failing runs and 68\% of GPT-5's, and it is invisible: the outcomes are confidently and consistently wrong, so completion-based and consistency-based monitoring both look healthy exactly when the agent should not be trusted. Lightweight pre-edit prompts do not close the gap. A second probe isolates the instinct to act: given an already-fixed bug, where the right move is to abstain, most models still edit the correct code. This \emph{action bias}, acting when no action is warranted, is exactly what completion metrics reward. The throughline is measurement: submit rate captures action, but trust requires validity. So evaluation must catch up: score agents by test-verified correctness over repeated runs, report its uncertainty, and reward those that know when not to act.
\end{abstract}

\section{Introduction}
\label{sec:intro}

How reliable are deployed coding agents? The answer depends entirely on \emph{what you measure}. By \emph{submit rate}, meaning whether the agent produces a patch, GPT-5 looks perfect: 100\% across 250 runs on SWE-bench~\citep{swebench}. By \emph{resolve rate}, whether the patch passes the project's tests, it drops to 44\%. The gap is not noise. On 19 of 50 tasks GPT-5 submits a confident-looking patch on \emph{every} run and \emph{every} patch fails. An operator watching submit rate sees flawless completion; only test execution reveals the patches are wrong.

This is, at heart, a measurement problem. Submit rate is a cheap, always-available \emph{indicator}, and the field quietly treats it as a stand-in for what we actually care about: the agent's ability to fix the bug. We show this stand-in has poor \emph{construct validity}. It mismeasures ability, and not evenly: the error is concentrated in one specific, dangerous, and \emph{invisible} failure mode.

We borrow the language of measurement science and psychometrics on purpose~\citep{messick1989validity,jacobs2021measurement}. An evaluation is \emph{reliable} if it gives the same score when repeated, and \emph{valid} if that score reflects what we mean to measure. The two are independent: a test can be perfectly reliable yet completely invalid. Our central object, \emph{silent semantic failure}, is exactly this case: an agent that reliably (5/5 runs) submits a patch that is always (0/5) wrong. It is \emph{silent} (invisible to submit- or completion-based monitoring: no crash, no empty patch, the run looks done) and \emph{semantic} (the patch is well-formed yet wrong in meaning). It is the most dangerous failure mode: confidently and consistently wrong, so reliability reads as competence with nothing to warn you.

Agent scoring also breaks the i.i.d.\ assumption behind single-run leaderboards: a trajectory is a chain of dependent steps, so a model's per-task outcome is a random variable, and one run is a noisy estimate. Measuring agents well means \emph{estimating} it from repeats, with uncertainty, not reading it off once.

Underneath both halves of our study is one tendency, \emph{action bias}: agents lean toward \emph{acting} rather than being right. When a bug exists, an over-eager agent submits a confident wrong patch (silent semantic failure); when none exists, it edits already-correct code instead of abstaining. Submit rate rewards the first; our specificity probe exposes the second.

We study these issues empirically across four frontier models on 50 SWE-bench Verified tasks spanning five repositories, five runs each, plus 750 guard-intervention trajectories. Our contributions are:

\begin{enumerate}
    \item \textbf{A measurement framework} (Section~\ref{sec:framework}) that casts coding-agent evaluation as construct measurement (construct, indicator, criterion) and organizes outcomes into a reliability~$\times$~validity grid in which \emph{silent semantic failure} is the reliable-but-invalid cell.

    \item \textbf{Convergent-validity failure of submit rate.} Submit rate and test-verified resolve rate disagree on model ranking. GPT-5 submits 100\%/resolves 44\%; Llama~4 submits 99\%/resolves 18\%; and although Gemini submits least often (70\%), it resolves more tasks than GPT-5 (50\% versus 44\%). Bootstrap leaderboards reverse the GPT-5/Gemini ranking depending on the indicator.

    \item \textbf{A failure taxonomy grounded in resolve rate.} Silent semantic failure (5/5 submit, 0/5 resolve) affects 33/50 Llama~4 tasks (SSFR~80\%) and 19/50 GPT-5 tasks (68\%), but only 4/50 Gemini tasks (16\%). Gemini fails \emph{honestly}, by abstaining.

    \item \textbf{Reliability is multi-faceted, stable, but not validity.} We measure consistency at three levels (pacing via step CV, procedure via command similarity, product via patch similarity) and find they are near-orthogonal ($r\le0.14$), so step counts alone are insufficient. The hierarchy holds across all five repositories, yet consistency only amplifies whatever interpretation the model commits to.

    \item \textbf{A mechanism for silent failure.} Within a model, silent-failure patches are far more self-similar than variable-failure patches: the agent re-derives the \emph{same} wrong fix (convergent misinterpretation, not random error). Across models, silent failures cluster on shared ``interpretation-trap'' tasks (8 tasks resolved by no model).

    \item \textbf{An abstention (specificity) probe.} Pre-applying the gold patch so the bug is already fixed, we measure whether agents recognize there is nothing to do. Models edit already-correct code (false action) on a large fraction of tasks, the specificity mirror of the same validity failure.

    \item \textbf{Process measures and a null intervention.} Two cheap trajectory signals (3-gram loop detection; trajectory length, $r=-0.26$) partially recover the missing validity signal without running tests, while two pre-edit guard interventions fail to improve resolve rate. Interpretation failures resist lightweight prompting.
\end{enumerate}

\section{A Measurement Framework for Coding Agents}
\label{sec:framework}

\paragraph{Construct, indicator, criterion.} We treat \emph{coding-agent capability on a task} as a hidden quantity $\theta$: how likely the agent is to produce a patch that actually fixes the issue. We never see $\theta$ directly, only indicators of it. The cheapest, the one a quick evaluation or monitoring dashboard reaches for, is the \textbf{submit rate}: did the agent return a non-empty patch (\texttt{exit\_status == "Submitted"})? Our \textbf{criterion}, the closest thing to ground truth, is the \textbf{resolve rate}: does the patch pass the project's full test suite under the SWE-bench harness~\citep{swebench_verified}? Construct validity asks whether the indicator tracks the criterion. It does not.

\paragraph{Reliability vs.\ validity.} An indicator is \emph{reliable} if repeated runs give the same value, and \emph{valid} if that value reflects $\theta$~\citep{messick1989validity}. The two are separate. Since we run each (model, task) pair five times, we can place every outcome on a reliability~$\times$~validity grid (populated per model in Figure~\ref{fig:relval_grid}, Appendix~\ref{app:consistency}). The cell that matters most is \textbf{reliable-but-invalid}: the agent returns the same confident, wrong patch every run. That is \emph{silent semantic failure}, and it is dangerous precisely because we usually read reliability as a good sign.

\paragraph{Scores are noisy draws, not fixed numbers.} A trajectory's steps are dependent, so each task's outcome is a random variable, not a fixed score. We therefore estimate its resolve probability $p_{\text{task}}$ from $k$ repeats with bootstrap uncertainty; a single-run leaderboard reports one draw and mistakes noise for capability. We quantify that noise with the coefficient of variation (CV) of step counts; throughout, CV is our reliability measure and resolve rate our validity criterion.

\section{Related Work}
\label{sec:related}

\paragraph{Measurement and benchmark validity.} A growing line of work treats AI benchmarks as measurement instruments: \citet{jacobs2021measurement} import construct-validity theory into ML, \citet{raji2021everything} critique general-ability claims, \citet{burnell2023rethinking} and \citet{liang2023holistic} push instance-level, multi-metric reporting, and \citet{kapoor2024ai} argue agent evaluation should prioritize deployment-relevant metrics. We operationalize this for coding agents: we name a concrete validity failure (the submit-rate trap), quantify it across models, and tie it to the reliability/validity distinction~\citep{messick1989validity}.

\paragraph{Agent benchmarks.} SWE-bench~\citep{swebench} and SWE-bench Verified~\citep{swebench_verified} evaluate agents on real GitHub issues; WebArena~\citep{webarena} and OSWorld~\citep{osworld} target web and computer use. Most report single-run aggregates, leaving reliability unexamined.

\paragraph{Consistency, agents, and safety.} Variance under repetition is documented for reasoning and QA~\citep{wang2023selfconsistency,elazar2021consistency,anon2026agents}; self-consistency~\citep{wang2023selfconsistency} exploits it via voting but assumes independent samples, which breaks for path-dependent trajectories. Agent-failure studies cover error categorization~\citep{agent_errors}, recovery~\citep{reflexion}, tool misuse~\citep{ruan2024toolemu}, and planning limits~\citep{valmeekam2023planning}. We characterize \emph{when} failures are reliable yet invalid: confident, consistent, and silent.

\section{Experimental Setup}
\label{sec:setup}

\paragraph{Benchmark and tasks.} We use SWE-bench Verified, selecting 50 tasks (10 per repository) across five mature Python projects: django (web), sympy (symbolic math), scikit-learn (ML), matplotlib (visualization), and astropy (astronomy), chosen for diversity in domain, size, and bug type.

\paragraph{Models.} We compare four models across capability and cost tiers: \textbf{Claude 4.5 Sonnet} (\texttt{claude-sonnet-4-5}), \textbf{GPT-5} (\texttt{openai-gpt-5}), \textbf{Gemini 3.1 Pro} (\texttt{gemini-3.1-pro}), and \textbf{Llama 4 Maverick} (\texttt{llama4-maverick}). All use hosted APIs with identical tools, system prompts, and task descriptions. Hosted snapshots can drift, so we pin the identifiers above and ran everything in a single 2026 window; the lack of immutable snapshot hashes is an acknowledged limit.

\paragraph{Agent framework and protocol.} We use mini-SWE-agent, a minimal bash-interface scaffold; each action executes in an isolated Docker container with fresh repository state. For each model--task pair we run 5 independent trials at temperature 0.5, max 250 steps. This yields 250 runs per model (1{,}000 baseline trajectories), plus 750 guard trajectories (below).

\paragraph{Metrics.} \textbf{Submit rate} is the fraction of runs producing a non-empty submitted patch (the indicator). \textbf{Resolve rate} is the fraction of runs whose patch passes all tests under the SWE-bench harness~\citep{swebench_verified} (the criterion). \textbf{Coefficient of variation} of step counts, $\text{CV} = \sigma_{\text{steps}}/\mu_{\text{steps}} \times 100\%$, is our reliability measure; lower is more consistent.

\paragraph{Guard interventions.} To test whether validity can be repaired by pre-edit prompting, we inject two guards before the agent's first code edit: an \textbf{interpretation guard} (state your understanding of the bug and fix strategy first; run on Claude and GPT-5, 500 trajectories) and a \textbf{test-anchored guard} (write and run a minimal failing test first; run on GPT-5, 250 trajectories).

\section{Results}
\label{sec:results}

\begin{table}[t]
\centering
\caption{Main results across 50 tasks and 5 repositories.}
\label{tab:main-results}
\begin{tabular}{lrrrr}
\toprule
Metric & Claude & GPT-5 & Llama 4 & Gemini \\
\midrule
Tasks & 50 & 50 & 50 & 50 \\
Total Runs & 250 & 250 & 250 & 250 \\
Mean CV (\%) & 17.7 & 30.3 & 58.1 & 44.9 \\
Median CV (\%) & 16.3 & 30.2 & 49.9 & 33.0 \\
Mean Steps & 53.3 & 10.5 & 28.6 & 78.5 \\
Submit Rate & 0.97 & 1.00 & 0.99 & 0.70 \\
Resolve Rate & 0.65 & 0.44 & 0.18 & 0.50 \\
\bottomrule
\end{tabular}
\end{table}

\begin{figure}[t]
\centering
\includegraphics[width=\textwidth]{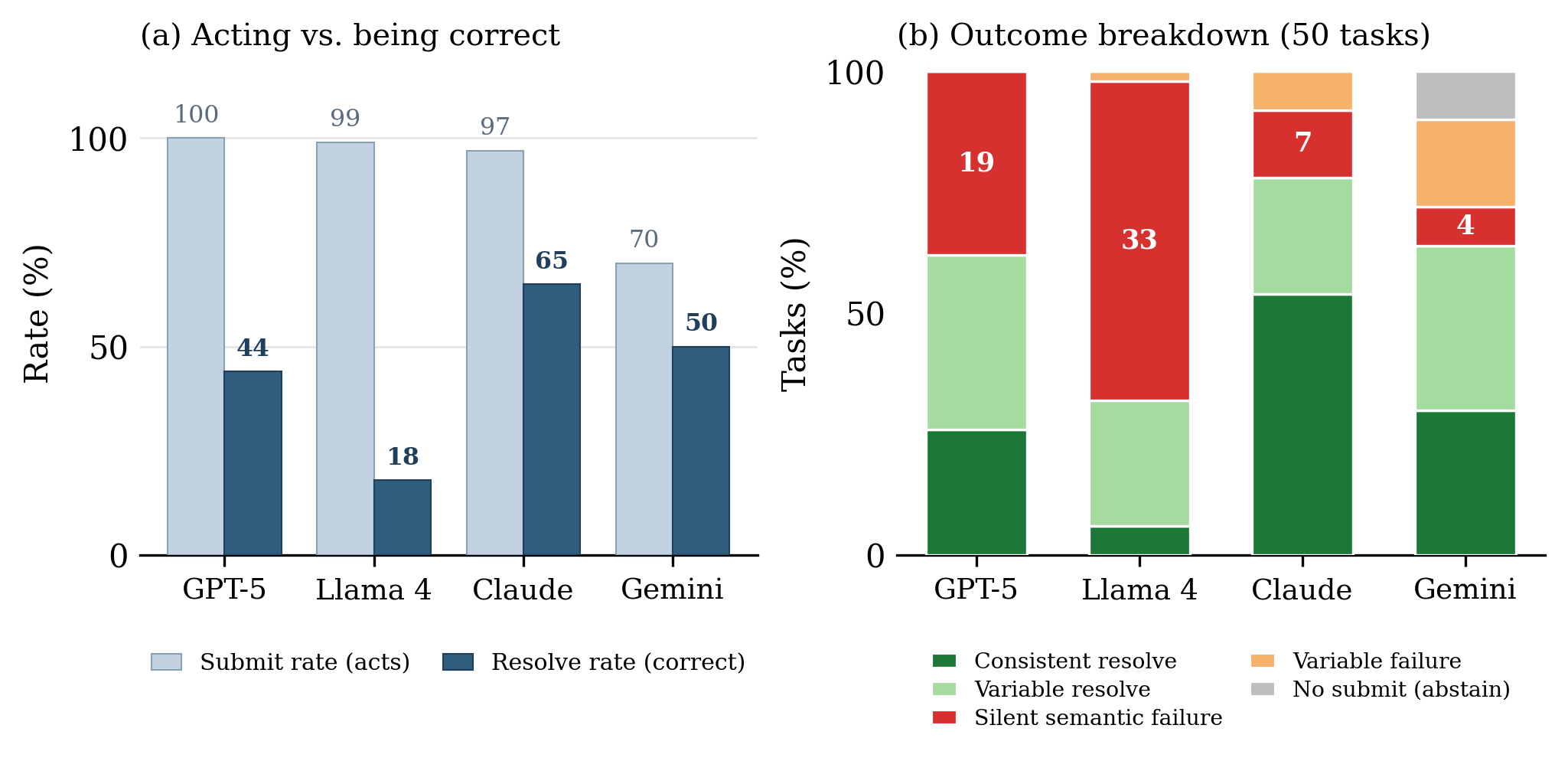}
\caption{The submit-rate trap at a glance. \textbf{(a)} Every model submits far more often than it resolves (bar labels are percentages); even though Gemini submits least often, it resolves more tasks than GPT-5 (50\% vs.\ 44\%). \textbf{(b)} Outcome breakdown over 50 tasks: \emph{silent semantic failure} (red; 5/5 submit, 0/5 resolve) dominates Llama~4 (33 tasks) and GPT-5 (19).}
\label{fig:headline}
\end{figure}

\subsection{Reliability: a stable behavioral-consistency hierarchy}
\label{sec:rq1}

Table~\ref{tab:main-results} shows a clear reliability hierarchy: Claude (CV 17.7\%) is the most consistent, then GPT-5 (30.3\%), Gemini (44.9\%), and Llama~4 (58.1\%). All pairwise differences are significant after Bonferroni correction except GPT-5 vs.\ Gemini and Gemini vs.\ Llama~4 (Appendix~\ref{app:stats}), and the ordering holds in all five repositories (Appendix~\ref{app:repo}). So reliability is a stable property of each model. But as we show next, being reliable does not make a measure valid.

\paragraph{Reliability is multi-faceted; step counts are only one facet.} One objection is that step-count CV is a thin measure of reliability: two runs can take the same number of steps yet run different commands and produce different patches. So we measure reliability at three levels (Table~\ref{tab:three-facet}): \emph{pacing} (step-count CV), \emph{procedure} (how similar the command sequences are across runs), and \emph{product} (how similar the submitted patches are). The three do not line up. Across all 200 (model, task) cells, step-CV barely correlates with command differences ($r{=}0.14$) and not at all with patch differences ($r{\approx}0$). Claude is the most consistent in \emph{pacing} but among the \emph{least} consistent in \emph{product} (patch similarity 0.13): it works at a steady rhythm but reaches the fix through genuinely different edits. Gemini is the reverse, with high patch similarity (0.42) driven partly by repeated command loops. The lesson: no single consistency number is enough, so we report all three and use them together below.

\begin{table}[t]
\centering
\caption{Reliability has three facets that do not coincide. \emph{Pacing} = step-count CV
(lower is more consistent); \emph{procedure} = mean cross-run command-sequence similarity;
\emph{product} = mean cross-run similarity of submitted patches (higher is more consistent).
Pooled across all 200 (model, task) cells, step-CV correlates only weakly with
command divergence ($r{=}0.14$) and essentially not with patch divergence ($r{\approx}0$): the facets are
near-orthogonal, so step counts alone do not capture procedural or product consistency.}
\label{tab:three-facet}
\vskip 0.05in
\begin{small}
\begin{tabular}{lrrr}
\toprule
 & Pacing & Procedure & Product \\
Model & Step-CV (\%) $\downarrow$ & Command sim. $\uparrow$ & Patch sim. $\uparrow$ \\
\midrule
Claude  & \textbf{17.7} & 0.13 & 0.13 \\
GPT-5   & 30.3 & 0.07 & 0.29 \\
Gemini  & 44.9 & \textbf{0.32} & \textbf{0.42} \\
Llama~4 & 58.1 & 0.12 & 0.31 \\
\bottomrule
\end{tabular}
\end{small}
\end{table}

\subsection{Validity: submit rate fails convergent validity}
\label{sec:rq2}

The indicator and the criterion disagree (Figure~\ref{fig:headline}a). GPT-5 submits 100\% but resolves 44\%; Llama~4 submits 99\% but resolves 18\%. Most tellingly, \textbf{Gemini submits the least often (70\%) but resolves more tasks than GPT-5 (50\% versus 44\%)}, second only to Claude: the indicator ranks Gemini last while the criterion ranks it second. A practitioner optimizing submit rate would pick the wrong model.

This convergent-validity failure is not just an artifact of point estimates. In bootstrap leaderboards (Appendices~\ref{app:stats} and~\ref{app:leaderboard}), submit rate ranks GPT-5 first in 98\% of resamples and Gemini last in 100\%; the resolve-rate bootstrap flips the GPT-5/Gemini order and ranks Claude first every time. The two metrics do not even agree on the podium. Claude and GPT-5 also look tied on submit rate (overlapping CIs) but are clearly apart on resolve rate, so a single-run submit comparison would call that gap real when it is not.

\paragraph{Is submit rate a straw man?} Rigorous leaderboards do report resolve rate, and we are not claiming otherwise. We have two targets. First, in \emph{deployment monitoring} there is rarely a trustworthy test for each fix, so operators watch completion signals (did the agent return a patch; did the run finish), which are submit-rate proxies. Second, the headline leaderboard number is single-attempt \emph{pass@1} (official SWE-bench, SWE-bench Pro, most agent scores), and a single run is high-variance; \emph{pass@$k$} is the multi-run estimate we recommend, but pass@1 still dominates. More broadly, submit rate interests us less as a flawed test surrogate than as a window onto an \emph{action bias} in evaluation: rewarding acting over being right.

\subsection{Reliable but invalid: the failure taxonomy}
\label{sec:taxonomy}

Using resolve rates, we classify each (model, task) pair by its 5-run submission/resolution pattern (Figure~\ref{fig:headline}b): \emph{consistent resolve} (5/5 resolved), \emph{variable resolve} (1--4/5), \emph{silent semantic failure} (5/5 submit, 0/5 resolve), \emph{variable failure} (partial submit, 0/5 resolve), and \emph{no-submit} (abstention; 0/5 submit). These map directly onto the reliability~$\times$~validity grid (Figure~\ref{fig:relval_grid}).

\begin{figure}[t]
\centering
\includegraphics[width=0.78\textwidth]{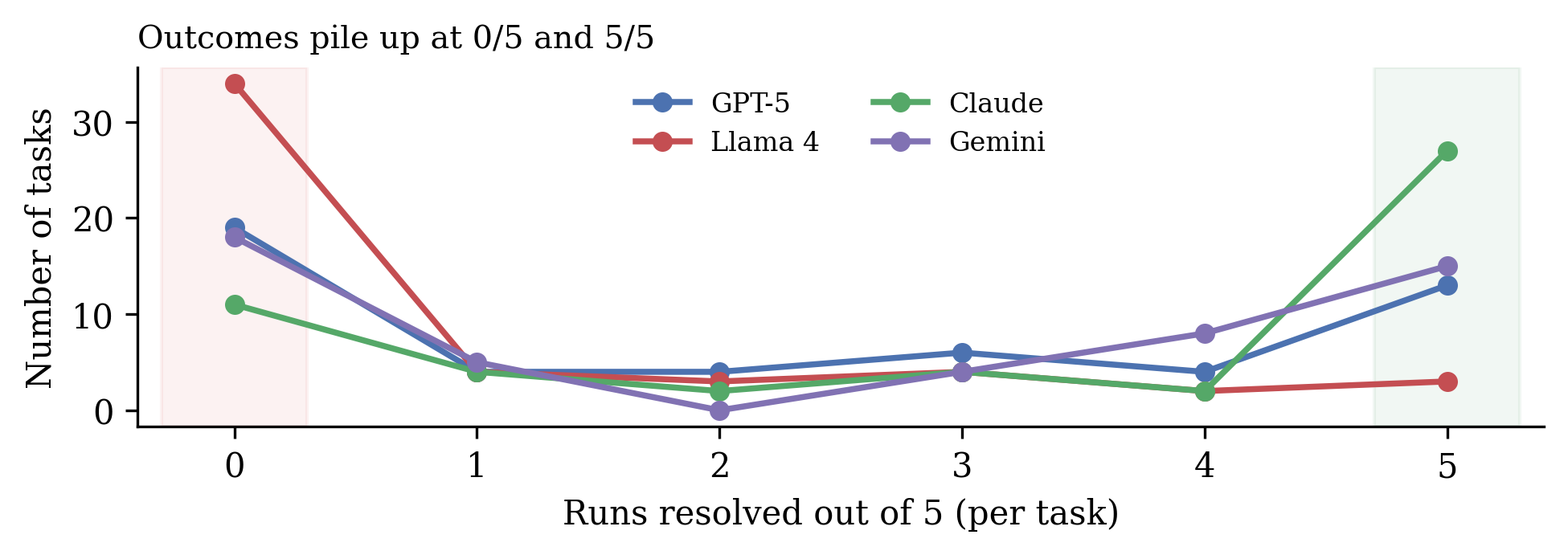}
\caption{Per-task resolve distribution: how many tasks (of 50) have $k$ of 5 runs resolve. Mass piles up at 0/5 and 5/5 for every model, sparse in between. Reliability amplifies whatever interpretation the model commits to: Claude peaks at 5/5, Llama~4 at 0/5.}
\label{fig:resolve_amp}
\end{figure}

Each model has a distinct profile (Figure~\ref{fig:headline}b). Claude is healthiest (27/50 consistent resolve, 7 SSF). GPT-5 is the most concerning of the high-submit models: \textbf{19/50 tasks are silent semantic failures}, submitting on all 5 and resolving none. Llama~4 is dominated by SSF (33/50). Gemini's failures are split between honest abstention (5 tasks at 0/5 submit) and variable outcomes. The per-task resolve distribution is sharply bimodal (Figure~\ref{fig:resolve_amp}): for almost every task either all five runs resolve or none do, so reliability amplifies whatever interpretation the model commits to rather than averaging it out.

We summarize the reliable-but-invalid cell with a \textbf{silent semantic failure rate} (SSFR): the fraction of non-resolving runs coming from tasks where all 5 runs submit but none resolve,
\begin{equation}
    \text{SSFR} = \frac{N_{\text{tasks: 5/5 submit, 0/5 resolve}} \times 5}{N_{\text{total non-resolving runs}}}.
\end{equation}
Llama~4's SSFR is \textbf{80\%} and GPT-5's \textbf{68\%}: over two-thirds of their failures are invisible to submission-based monitoring. Claude's is 40\%; Gemini's only 16\%. This is the paper's central measurement finding: \textbf{the most dangerous failure mode is not failing to act, but acting reliably and wrongly}, and submit-rate monitoring is blind to exactly this cell.

\subsection{Why do silent failures happen? Convergent misreadings and shared traps}
\label{sec:why}

Failing tests is no surprise; the measurement question is \emph{why} the failure is silent and reliable. Two findings from the trajectories explain it.

\textbf{(1) Within a model, the wrong patches are nearly identical.} If silent failure were random error, the five wrong patches for a task would differ. They do not. On three measures (textual diff overlap, code-token Jaccard, sentence-embedding cosine), SSF patches are far more alike than a cross-task baseline (Table~\ref{tab:ssf-why}): unrelated patches score near zero, while SSF patches score many times higher. The agent rebuilds the same wrong fix each run, a fixed misreading rather than bad luck, so reliability just reproduces the mistake.

\begin{table}[t]
\centering
\caption{Why silent failures happen: when an agent silently fails (5/5 submit, 0/5 resolve),
its five patches are far more similar to one another than \emph{unrelated} patches from
different tasks (cross-task baseline), on \emph{every} similarity measure: textual diff
overlap, code-token Jaccard, and sentence-embedding cosine. The agent converges on the
\emph{same} wrong fix rather than failing randomly. (GPT-5 has no variable-failure tasks: it
submits on every run, so all its non-resolving tasks are silent semantic failures.)}
\label{tab:ssf-why}
\vskip 0.05in
\begin{small}
\begin{tabular}{lrrrr}
\toprule
 & SSF & \multicolumn{3}{c}{Cross-run patch similarity (SSF $/$ cross-task baseline)} \\
\cmidrule(l){3-5}
Model & tasks & Textual diff & Token Jaccard & Embedding cosine \\
\midrule
Claude  & 7  & 0.18 / 0.01 & 0.41 / 0.10 & 0.85 / 0.48 \\
GPT-5   & 19 & 0.27 / 0.01 & 0.36 / 0.05 & 0.85 / 0.53 \\
Gemini  & 4  & 0.39 / 0.01 & 0.67 / 0.09 & 0.91 / 0.57 \\
Llama~4 & 33 & 0.26 / 0.01 & 0.38 / 0.05 & 0.80 / 0.52 \\
\bottomrule
\end{tabular}
\end{small}
\end{table}

\textbf{(2) Across models, silent failures cluster on the same tasks.} The SSF task-sets of GPT-5 and Llama~4 overlap heavily (Jaccard 0.49); 7 tasks are SSF for at least 3 of the 4 models, and 8 are solved by \emph{no} model on any run (Figure~\ref{fig:ssf_overlap}). These are \emph{interpretation traps}: the obvious reading of the issue points away from the real fix. But 16 tasks trap only one model, so silent failure has two sources, hard tasks and over-eager models, and good evaluation must vary both.

\subsection{Cheap process measures recover part of the missing signal}
\label{sec:detection}

Can we flag the reliable-but-invalid cell \emph{without} running tests? From the 1{,}000 baseline trajectories we extract a few cheap features. A \textbf{3-gram loop detector} (any run of 3 commands that repeats $\geq 3$ times) catches 100\% of Gemini's 0/5 no-submit (abstention) tasks, with no false positives on Claude or GPT-5; a step-ceiling rule ($>$245 steps) catches 80\%. \textbf{Trajectory length} also predicts silent semantic failure: across the 200 (model, task) cells, shorter average runs go with it ($r=-0.26$, $p<0.001$), and silent-failure cells have slightly higher phase-mix entropy ($r=+0.20$, $p<0.01$). Models that lock onto a wrong reading early produce short, confident, wrong runs. These are test-free signals that recover part of what submit rate throws away.

\subsection{Validity is not repaired by pre-edit prompting}
\label{sec:guard}

If invalid submissions stem from interpretation errors, can a pre-edit nudge fix them? We test two guards on GPT-5 (interpretation, test-anchored) and one on Claude (interpretation); full numbers in Appendix~\ref{app:guard} (Table~\ref{tab:guard-effect}).

Neither guard moves overall resolve rate: the interpretation guard leaves Claude at $65.2\%\to64.8\%$ and GPT-5 at $44.4\%\to46.0\%$; the test-anchored guard leaves GPT-5 at $44.4\%\to44.0\%$. The test-anchored guard helps and hurts unevenly across repositories, gaining on isolated bugs (astropy +4pp, django +6pp, matplotlib +6pp) but losing on complex suites (scikit-learn $-10$pp, sympy $-8$pp), because writing a test first (17.2 vs 10.5 steps) burns budget on hard tasks. It does cut GPT-5's SSF tasks from 19/50 to 16/50. Two very different prompts both failing moves the conclusion from ``one prompt didn't work'' to ``a whole class of pre-edit prompts doesn't repair validity.''

\section{Specificity: do agents know when to abstain?}
\label{sec:abstain}

Resolve rate measures \emph{sensitivity}: when a bug exists, does the agent fix it? A good
capability measure also needs \emph{specificity}: when there is nothing to fix, does the agent
notice and stop? We test this directly. For a subset of tasks across repositories, we apply and
commit the \emph{gold} patch before the agent starts, so the bug in the issue is
\textbf{already fixed}; the agent still gets the original bug report. A well-calibrated agent
should reproduce the issue, see it is resolved, and abstain (submit nothing). An over-confident
one will ``fix'' the non-bug anyway, editing code that is already correct. We label each run
\emph{false action} (submits a non-empty patch), \emph{clean abstain} (submits an empty patch),
or \emph{no submit} (runs out of steps without deciding).

\begin{figure}[t]
\centering
\includegraphics[width=0.7\textwidth]{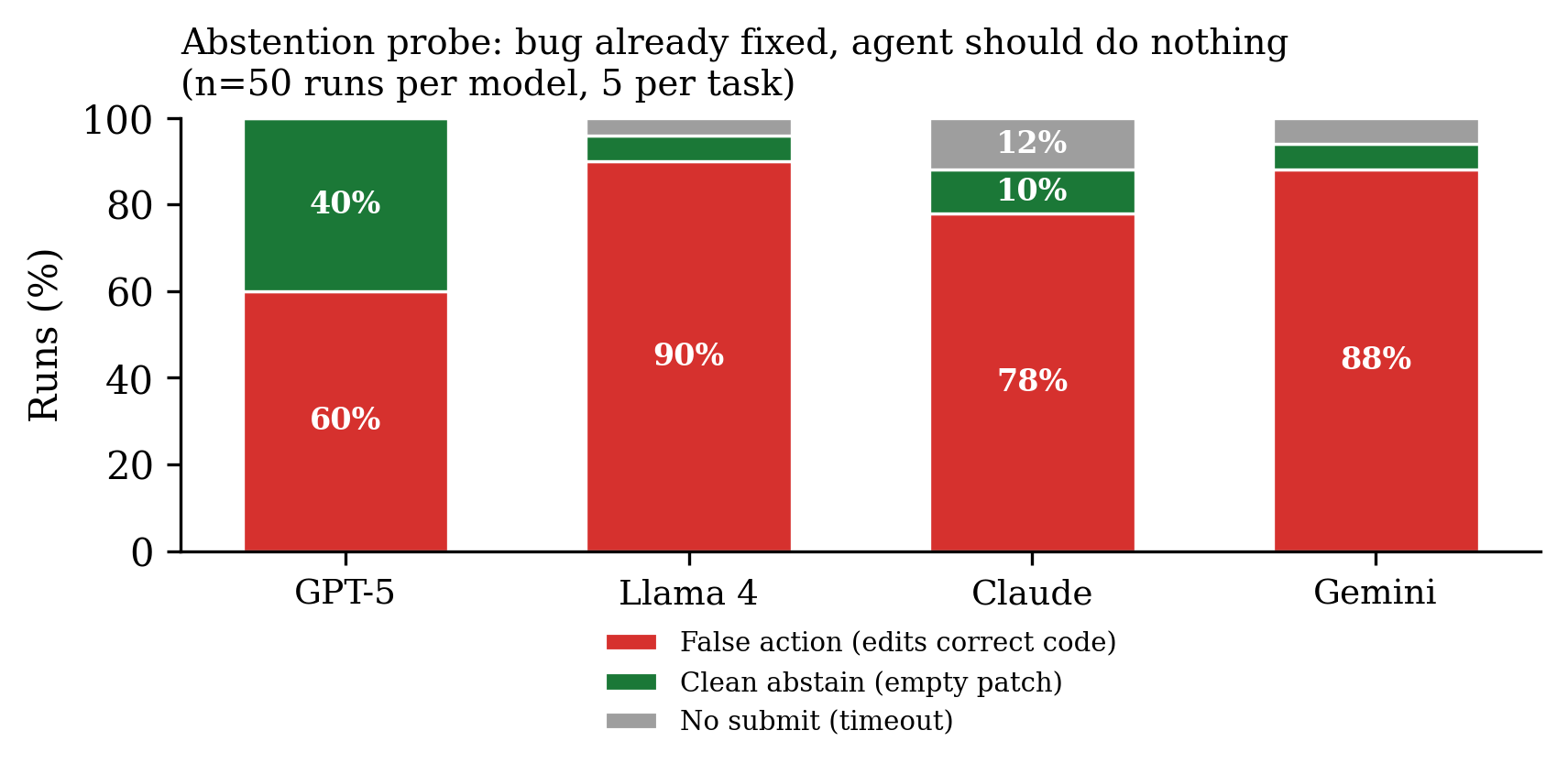}
\caption{Abstention probe: the gold patch is pre-applied so the bug is already fixed and the agent should do nothing. Per-model split into false action (a spurious edit to correct code), clean abstain (empty submit), and no submit (timeout). All four models edit correct code on a majority of runs ($n=50$ per model, five per task).}
\label{fig:abstention}
\end{figure}

\textbf{All four models routinely act on non-bugs.} On 60--90\% of runs ($n=50$ per model), every model edits correct code instead of abstaining (Figure~\ref{fig:abstention}); only GPT-5 abstains cleanly with any frequency (40\%), yet still over-acts on the majority. ``Fixing'' something that is not broken is the specificity side of the same validity failure: a metric that rewards \emph{acting} even when the right output is no change. Abstention is a real capability that submission-based evaluation punishes.

\section{Discussion: a measurement protocol}
\label{sec:discussion}

We turn the findings into a measurement protocol.

\begin{enumerate}
\itemsep1pt
\item \textbf{Score against a criterion, not an indicator}: report test-verified resolve rate; GPT-5's 100\%/44\% gap shows submit rate is far from sufficient.
\item \textbf{Estimate, don't read off}: run $k\geq3$ trials with bootstrap CIs. Best-of-3 raises Claude 65$\to$74\% and GPT-5 44$\to$58\% (Appendix~\ref{app:cost}).
\item \textbf{Report reliability beside accuracy}: high agreement with low resolve is the silent-failure danger sign.
\item \textbf{Add test-free process measures}: 3-gram loop detection (100\% recall, 0\% FP) and short-trajectory flags.
\item \textbf{Measure specificity too}: an agent that cannot abstain on already-fixed code is miscalibrated even at high resolve rate.
\end{enumerate}

Underlying all five: reliability helps only \emph{conditional on validity}; absent a criterion, high consistency just manufactures false confidence.

\section{Limitations}
\label{sec:limitations}

\textbf{Criterion validity.} Resolve rate is an imperfect criterion (SWE-bench suites can be incomplete) but far closer to the construct than submit rate. \textbf{Single benchmark and language.} All tasks are SWE-bench Verified \emph{Python}; compiled languages (Java, C++, Rust) add a compiler that may catch over-confident edits, so SSF rates there are an open question. \textbf{Single scaffold.} One minimal scaffold (mini-SWE-agent); richer scaffolds may shift rates, though our null guard suggests lightweight fixes are insufficient. \textbf{Scale.} 50 tasks is modest; we mitigate with paired tests and bootstrap CIs. \textbf{Abstention scope.} The specificity probe (Section~\ref{sec:abstain}) uses $n=50$ per model on a 10-task subset, not the full benchmark.

\paragraph{Future work.} Cross-language and cross-scaffold replication, structural interventions (multi-agent debate, self-verification) for the interpretation failures prompting could not fix, and scaling the abstention probe into a full specificity benchmark paired with resolve rate.

\section{Conclusion}
\label{sec:conclusion}

Submit rate measures whether a coding agent \emph{acts}, not whether it is \emph{right}, and the two diverge dangerously. It reverses the GPT-5/Gemini resolve ranking, and its errors concentrate in \emph{silent semantic failure} (SSFR up to 80\%): reliable, confident, test-failing patches where agents repeat the same wrong fix. As agents move into production, measurement must follow: test-verified, multi-run estimation reporting reliability \emph{and} specificity, not the submit-rate trap.

\bibliography{references}
\bibliographystyle{colm2026_conference}

\appendix

\FloatBarrier
\section{Case Studies}
\label{app:casestudy}

\paragraph{Silent semantic failure (astropy-13236).} Adding a structured numpy array to an Astropy Table silently converts it to \texttt{NdarrayMixin}; the fix is to remove the conversion (4 lines). Claude instead read the task as ``add a deprecation warning but preserve behavior,'' spending 30--50 steps per run on a \texttt{FutureWarning}. All 5 runs submitted; none resolved. GPT-5 made the same interpretation error in 7 steps (submit 5/5, resolve 0/5). This is the reliable-but-invalid cell in miniature: perfect submission, zero resolution, no behavioral variance to signal the problem.

\begin{figure}[h]
\centering
\includegraphics[width=0.62\textwidth]{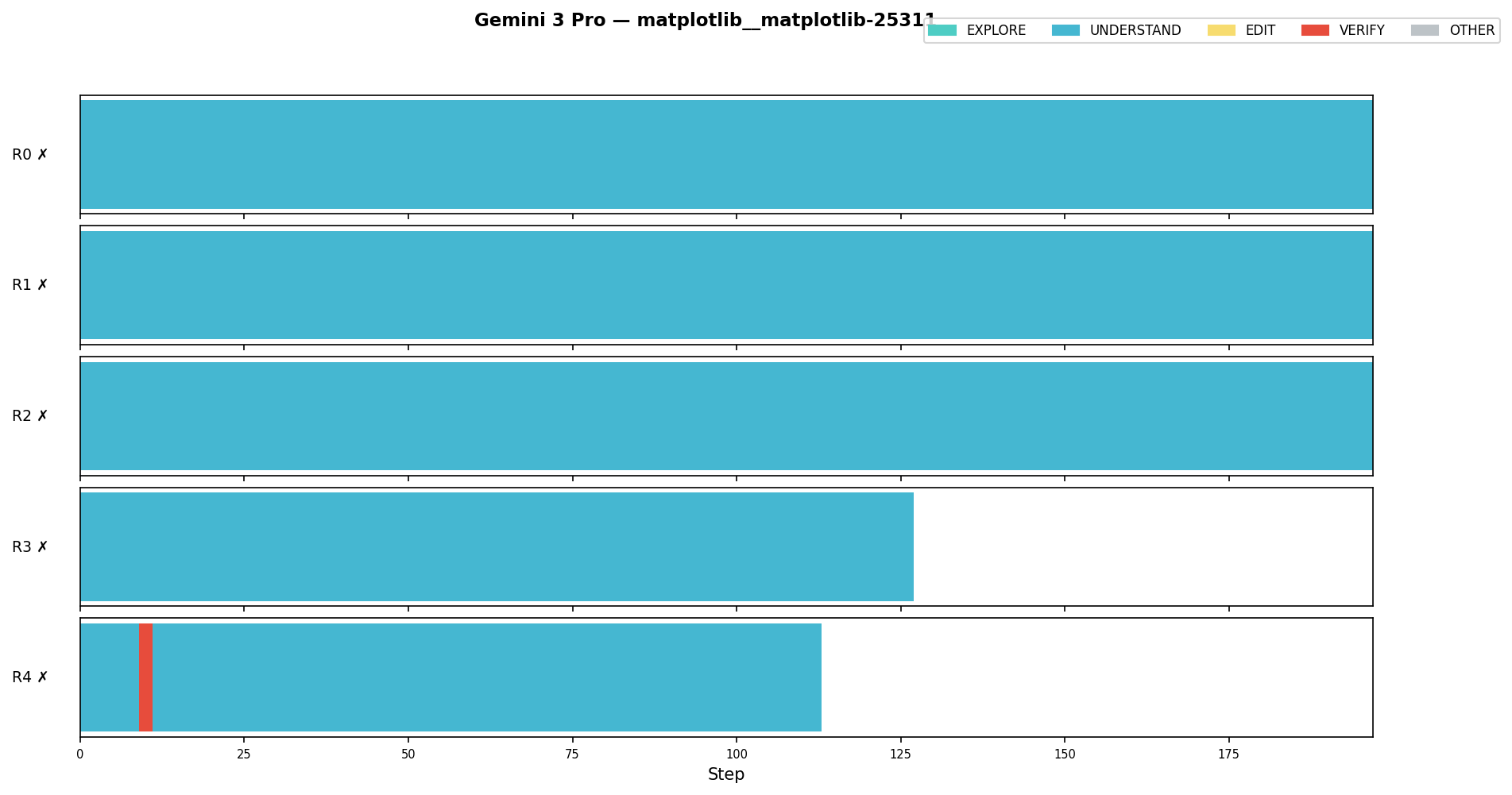}
\caption{Gemini trajectory timelines on matplotlib-25311 (one row per run; teal=explore, blue=understand, yellow=edit, red=verify). All five runs are degenerate read-loops with no edit phase.}
\label{fig:timeline}
\end{figure}

\paragraph{The read-loop trap (matplotlib-25311).} A draggable-legend figure cannot be pickled; the fix needs \texttt{\_\_getstate\_\_}/\texttt{\_\_setstate\_\_} on an offsetbox class. Claude (5/5, mean 74 steps) and GPT-5 (5/5, mean 10 steps) resolved it. Gemini submitted \textbf{0/5}, exhausting the 250-step limit on every run in a degenerate read loop, cycling the same \texttt{grep}/\texttt{sed} commands without ever editing (Figure~\ref{fig:timeline}). This failure is costly but \emph{honest}: 0/5 submit, no variance suggesting a retry would help, and caught perfectly by the 3-gram loop detector.

\FloatBarrier
\section{Guard Intervention Details}
\label{app:guard}
\begin{table}[t]
\centering
\caption{Effect of guard interventions on GPT-5 and Claude. Neither interpretation guard nor test-anchored guard significantly improves resolve rate.}
\label{tab:guard-effect}
\begin{footnotesize}
\begin{tabular}{lrrr}
\toprule
 & GPT-5 & GPT-5 & GPT-5 \\
Metric & Baseline & Interp.\ Guard & Test Guard \\
\midrule
Resolve Rate & 0.444 & 0.460 & 0.440 \\
Submit Rate & 1.000 & 0.996 & 0.976 \\
Mean Steps & 10.5 & 11.7 & 17.2 \\
SSF tasks & 19/50 & n/a & 16/50 \\
\midrule
 & Claude & Claude & \\
Metric & Baseline & Interp.\ Guard & \\
\midrule
Resolve Rate & 0.652 & 0.648 & n/a \\
Submit Rate & 0.968 & 0.972 & n/a \\
Mean Steps & 53.3 & 50.3 & n/a \\
\bottomrule
\end{tabular}
\end{footnotesize}
\end{table}

\FloatBarrier
\section{Cross-Repository Reliability}
\label{app:repo}
\begin{table}[t]
\centering
\caption{Mean CV (\%) by repository and model.}
\label{tab:per-repo}
\begin{tabular}{lrrrr}
\toprule
Repository & Claude & GPT-5 & Llama 4 & Gemini \\
\midrule
django & 17.8 & 26.6 & 54.0 & 46.4 \\
sympy & 23.4 & 29.3 & 64.7 & 46.8 \\
scikit-learn & 18.0 & 26.8 & 67.9 & 50.4 \\
matplotlib & 13.9 & 36.7 & 44.4 & 35.4 \\
astropy & 15.2 & 32.2 & 59.8 & 45.5 \\
\bottomrule
\end{tabular}
\end{table}
\begin{figure}[h]
\centering
\includegraphics[width=0.6\textwidth]{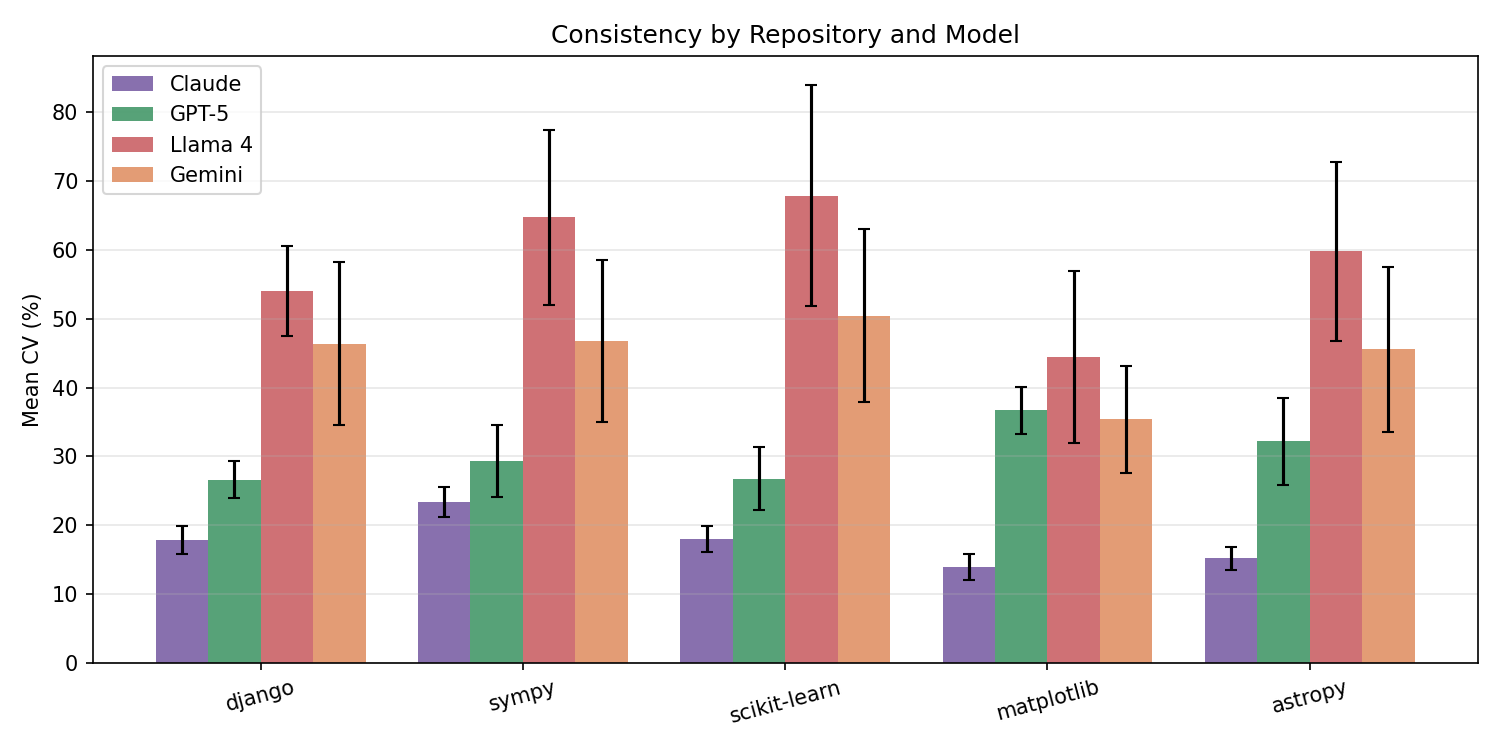}
\caption{Mean CV by repository and model; the reliability hierarchy holds across all five repositories.}
\label{fig:repo}
\end{figure}

\begin{figure}[h]
\centering
\includegraphics[width=0.6\textwidth]{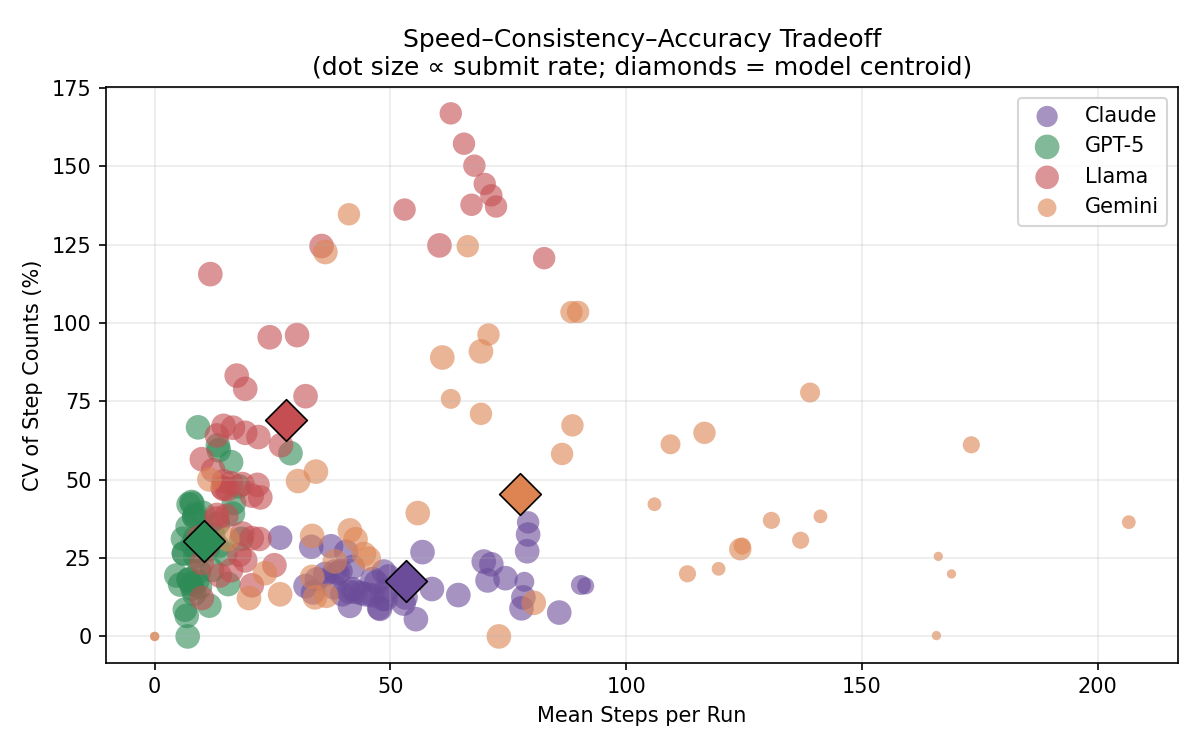}
\caption{Steps vs.\ behavioral consistency (CV); each point is one task. Claude clusters high-step/low-CV; GPT-5 is fast with moderate variance; Llama spans high variance; Gemini ranges widely. Speed and reliability do not imply validity.}
\label{fig:tradeoff}
\end{figure}

\FloatBarrier
\section{Supplementary Consistency Figures}
\label{app:consistency}

\begin{figure}[h]
\centering
\includegraphics[width=\textwidth]{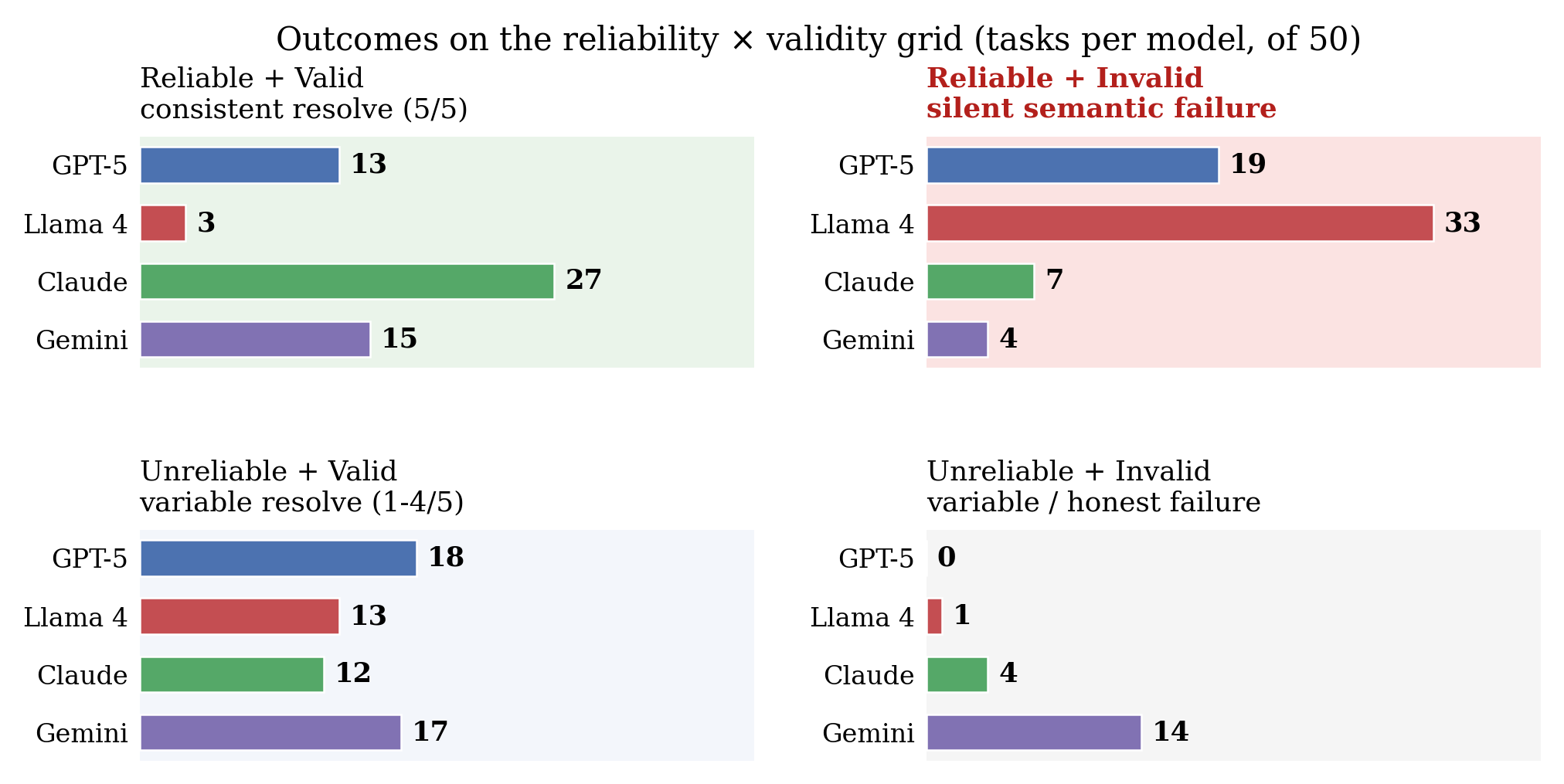}
\caption{The reliability~$\times$~validity grid populated with task counts per model (of 50). Each model's mass sits in a different quadrant: Claude in reliable-and-valid (27 consistent resolves), Llama~4 and GPT-5 in the reliable-but-invalid \emph{silent semantic failure} cell (33 and 19), and Gemini spread toward the honest-failure quadrant.}
\label{fig:relval_grid}
\end{figure}

\begin{figure}[h]
\centering
\includegraphics[width=0.9\textwidth]{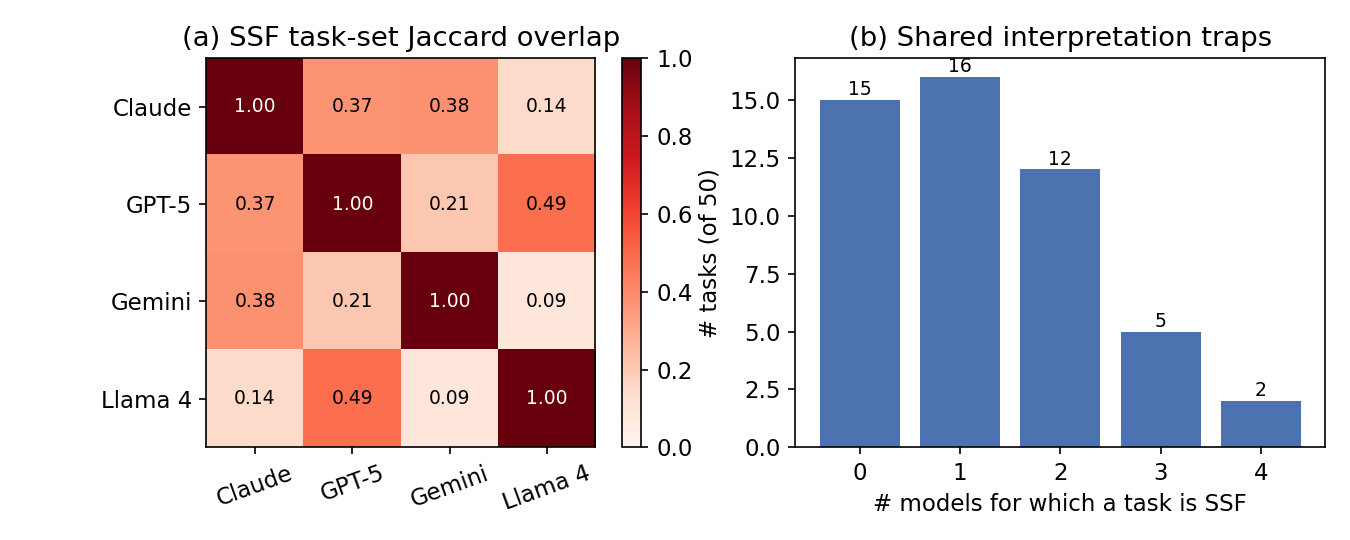}
\caption{Cross-model silent-semantic-failure overlap. (a) Jaccard overlap of SSF task-sets: GPT-5 and Llama~4 share most. (b) Number of tasks that are SSF for exactly $k$ models; 7 tasks trap $\geq3$ models (shared interpretation traps), while many are model-specific.}
\label{fig:ssf_overlap}
\end{figure}

\begin{figure}[h]
\centering
\includegraphics[width=0.55\textwidth]{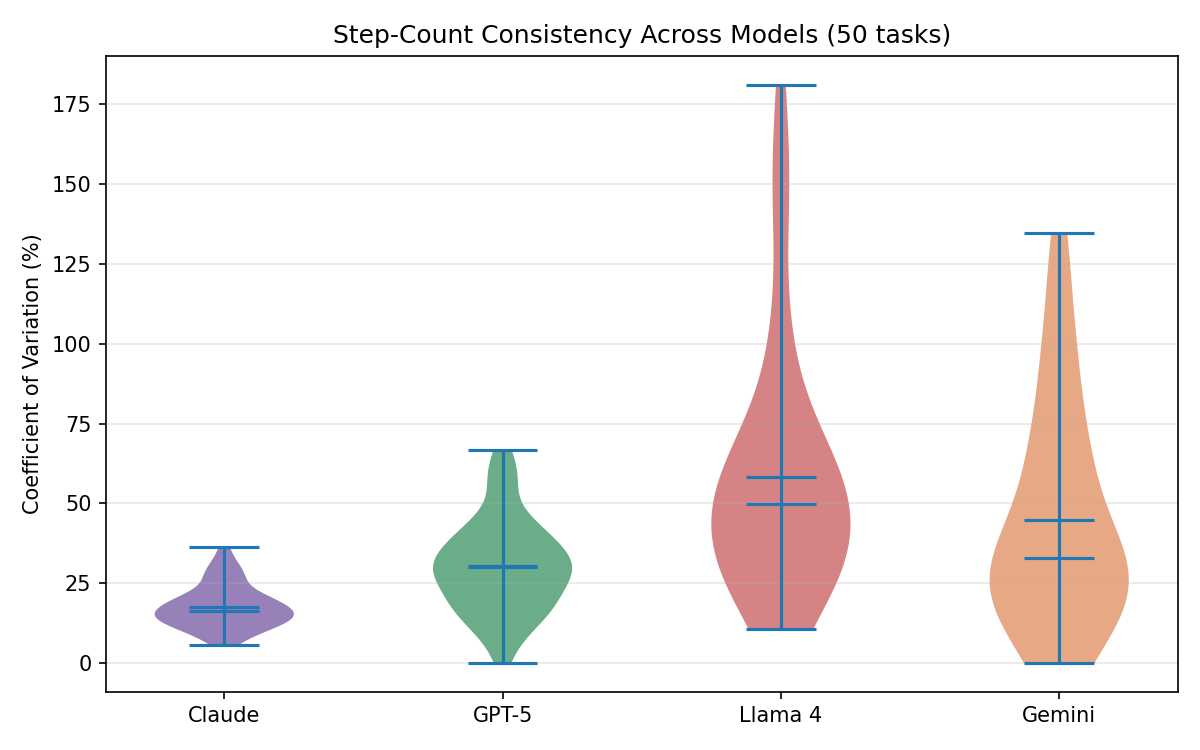}
\caption{Per-task behavioral consistency (CV) across four models (50 tasks each). The gap between Claude and the others is visually stark.}
\label{fig:cv_dist}
\end{figure}

\begin{figure}[h]
\centering
\includegraphics[width=0.95\textwidth]{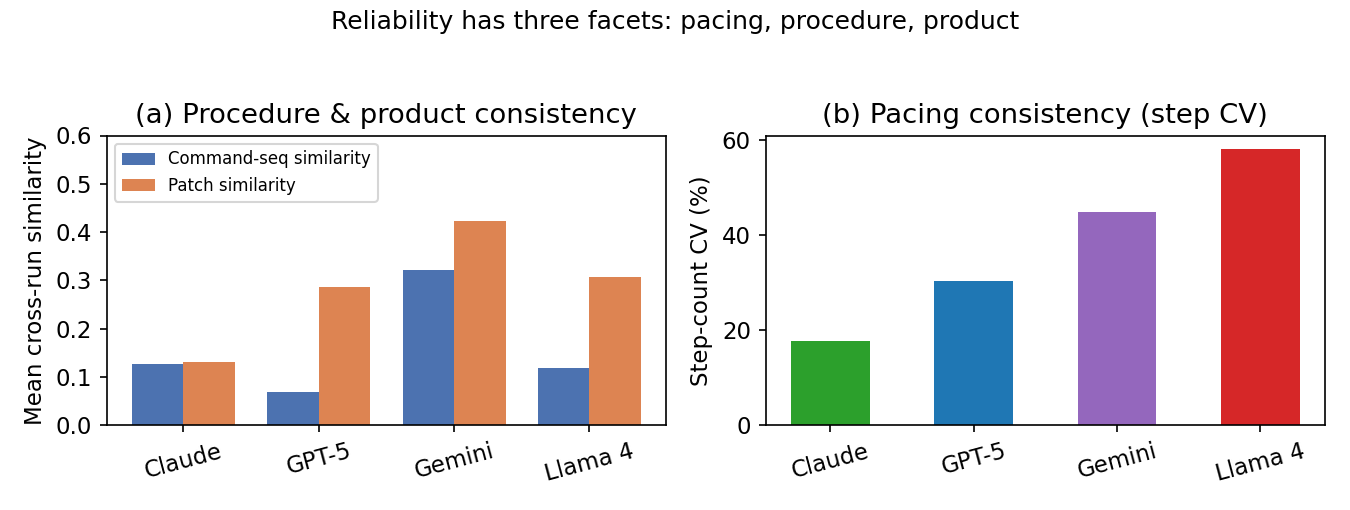}
\caption{The three facets of reliability per model. (a) procedure (command-sequence) and product (patch) similarity; (b) pacing (step CV). The facets do not coincide.}
\label{fig:three_facet}
\end{figure}

\begin{figure}[h]
\centering
\includegraphics[width=0.55\textwidth]{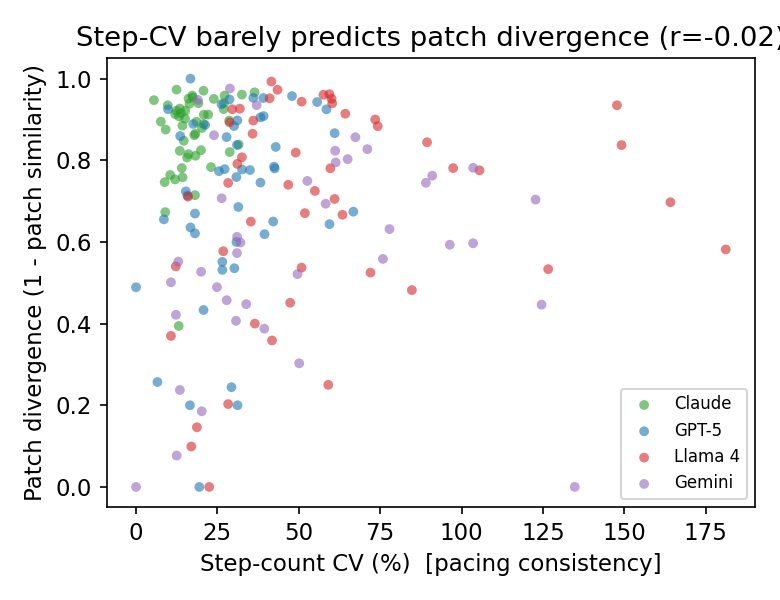}
\caption{Per-task step-CV vs.\ patch divergence (pooled, colored by model). The near-zero correlation shows step-count consistency does not capture patch-level consistency.}
\label{fig:cv_vs_patch}
\end{figure}

\begin{figure}[h]
\centering
\includegraphics[width=0.7\textwidth]{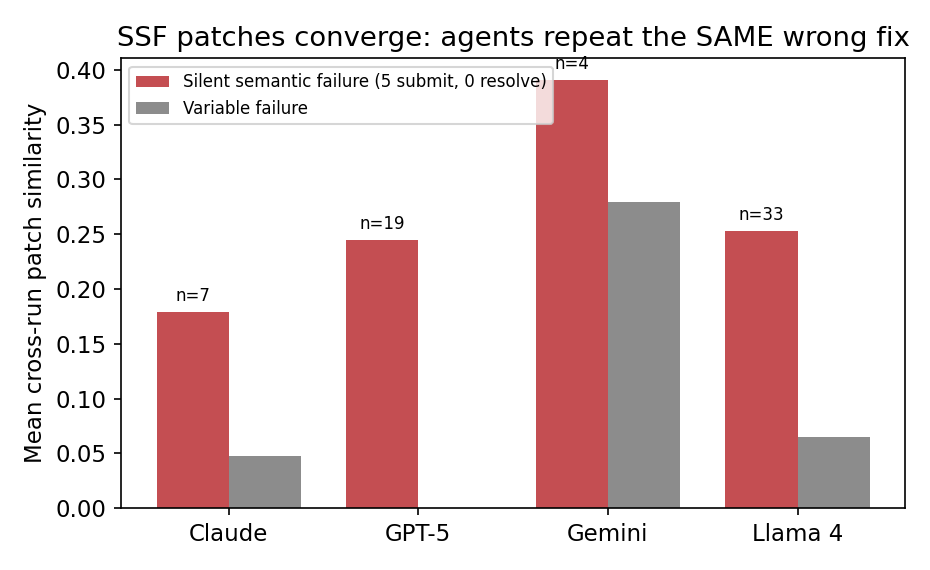}
\caption{Cross-run patch similarity for silent-semantic-failure tasks vs.\ variable-failure tasks. SSF patches are far more self-similar: agents converge on the same wrong fix.}
\label{fig:ssf_conv}
\end{figure}

\FloatBarrier
\section{Statistical Details}
\label{app:stats}

\paragraph{Pairwise CV comparisons (paired $t$-tests on per-task CV, 50 shared tasks; Bonferroni $\alpha=0.0083$):}
Claude vs.\ GPT-5 $\Delta=-12.6$\%, $p<0.001$; Claude vs.\ Gemini $\Delta=-27.2$\%, $p<0.001$; Claude vs.\ Llama~4 $\Delta=-40.4$\%, $p<0.001$; GPT-5 vs.\ Gemini $\Delta=-14.6$\%, $p=0.009$ (borderline); GPT-5 vs.\ Llama~4 $\Delta=-27.8$\%, $p<0.001$; Gemini vs.\ Llama~4 $\Delta=-13.2$\%, $p=0.02$ (n.s.\ after correction).

\paragraph{Submit-rate bootstrap ($n=1000$):} Claude 0.969 [0.936, 0.996] (rank-1 in 1.7\%); GPT-5 1.000 [1.000, 1.000] (98.3\%); Gemini 0.697 [0.600, 0.796] (0\%); Llama~4 0.988 [0.972, 1.000] (0\%).

\paragraph{Resolve-rate bootstrap ($n=1000$):} Claude 0.651 [0.600, 0.700] (rank-1 in all samples); Gemini 0.498 [0.420, 0.560]; GPT-5 0.444 [0.360, 0.520]; Llama~4 0.180 [0.100, 0.260]. The resolve-rate bootstrap \emph{reverses} the GPT-5 vs.\ Gemini ranking relative to submit rate.

\FloatBarrier
\section{Cost-Benefit of Multi-Run Evaluation}
\label{app:cost}
\begin{table}[t]
\centering
\caption{Cost-benefit analysis of multi-run evaluation strategies on \emph{submit rate}. Cost is mean actual API cost per task. Majority vote uses hard majority; Best-of-$k$ is the oracle upper bound (task counts as submitted if \emph{any} of $k$ runs submits). Resolve-rate-based best-of-$k$ values appear in the recommendations text.}
\label{tab:cost_benefit}
\resizebox{\columnwidth}{!}{
\begin{tabular}{llrrrr}
\toprule
Model & Strategy & Submit & Cost (\$) & Rel.\ Cost & Gain/\$\\
\midrule
Claude & Single \textit{(k=1)} & 0.968 & 3.26 & 1.0x & n/a\\
Claude & Majority \textit{(k=3)} & 0.974 & 9.78 & 3.0x & 0.001\\
Claude & Majority \textit{(k=5)} & 0.980 & 16.31 & 5.0x & 0.001\\
Claude & Best-of-3 & 1.000 & 9.81 & 3.0x & 0.003\\
Claude & Best-of-5 & 1.000 & 16.31 & 5.0x & 0.002\\
\midrule
GPT-5 & Single \textit{(k=1)} & 1.000 & 0.62 & 1.0x & n/a\\
GPT-5 & Majority \textit{(k=3)} & 1.000 & 1.85 & 3.0x & 0.000\\
GPT-5 & Best-of-5 & 1.000 & 3.08 & 5.0x & 0.000\\
\midrule
Llama~4 & Single \textit{(k=1)} & 0.988 & 0.34 & 1.0x & n/a\\
Llama~4 & Majority \textit{(k=3)} & 1.000 & 1.03 & 3.0x & 0.012\\
Llama~4 & Best-of-3 & 1.000 & 1.09 & 3.2x & 0.011\\
\midrule
Gemini & Single \textit{(k=1)} & 0.696 & 3.56 & 1.0x & n/a\\
Gemini & Majority \textit{(k=3)} & 0.712 & 10.68 & 3.0x & 0.001\\
Gemini & Best-of-3 & 0.840 & 10.32 & 2.9x & 0.014\\
Gemini & Best-of-5 & 0.900 & 17.80 & 5.0x & 0.011\\
\bottomrule
\end{tabular}
}
\end{table}

\FloatBarrier
\section{Leaderboard Bootstrap}
\label{app:leaderboard}
\begin{table}[t]
\centering
\caption{Leaderboard contextualization via bootstrap resampling (1000 samples) on \emph{submit rate}. Rank-$k$ is the fraction of bootstrap samples in which the model achieved that rank. Single-run reliability is the fraction of runs agreeing with the 5-run majority vote.}
\label{tab:leaderboard_ci}
\resizebox{\columnwidth}{!}{
\begin{tabular}{lrrrrr}
\toprule
Model & Mean Submit & 95\% CI & Rank-1 & Rank-2 & Reliability\\
\midrule
Claude & 0.969 & [0.936, 0.996] & 1.7\% & 14.5\% & 0.972\\
GPT-5 & 1.000 & [1.000, 1.000] & 98.3\% & 1.7\% & 1.000\\
Llama~4 & 0.988 & [0.972, 1.000] & 0.0\% & 83.8\% & 0.988\\
Gemini & 0.697 & [0.600, 0.796] & 0.0\% & 0.0\% & 0.876\\
\bottomrule
\end{tabular}
}
\end{table}

\end{document}